\documentstyle{article}
\begin{document}

\centerline{ \Large \bf Negative moments of characteristic
polynomials of random} \centerline{ \Large \bf GOE matrices and
singularity-dominated strong fluctuations}

\centerline{}

\vskip 0.4cm \centerline{\large \bf Yan V. Fyodorov$^1$ and
Jonathan P. Keating$^2$}

\vskip 0.3cm

\centerline{$^1$Department of Mathematical Sciences, Brunel
University} \centerline{Uxbridge, UB8 3PH, UK} \centerline{$^2$
School of Mathematics, University of Bristol} \centerline{Bristol,
BS8 1TW, UK} \vskip 0.3cm

\begin{abstract}
We calculate the negative integer moments of the (regularized)
characteristic polynomials of $N\times N$ random matrices taken
from the Gaussian Orthogonal Ensemble (GOE) in the limit as $N
\rightarrow \infty$. The results agree nontrivially with a recent
conjecture of Berry \& Keating motivated by techniques developed
in the theory of singularity-dominated strong fluctuations.  This
is the first example where nontrivial predictions obtained using
these techniques have been proved.

\end{abstract}

\section{Introduction}

Let $\hat{H}=\hat{H}^T$ (we here use the symbol $^T$ to denote
matrix or vector transposition and $^*$ to denote complex
conjugation) be an $N\times N$ random symmetric matrix with real
entries distributed according to the standard joint probability
density of the Gaussian Orthogonal Ensemble (GOE) of random matrix
theory --
\begin{equation} \label{GOE}
{\cal P}(\hat{H})=C_Ne^{-\frac{N}{2J^2}\mbox{Tr}\hat{H}^2},
\end{equation}
with respect to the measure $d\hat{H}=\prod_{i=1}^N
dH_{ii}\prod_{i<j}dH_{ij}$, where the normalization constant $C_N$
is given by
\begin{equation}\label{norm}
C_N=\frac{1}{2^{N/2}}\left(\frac{N}{\pi J^2}\right)^
{\frac{N(N+1)}{4}}
\end{equation}
 -- and let
\begin{equation} \label{Z}
Z_N(\mu)=\det\left(\mu {\bf 1}_N-\hat{H}\right)
\end{equation}
denote its characteristic polynomial.  We shall here be interested
in the negative integer moments of $|Z|$, defined by averaging
over the GOE, when $\mbox{Im}\mu>0$, in the limit as $N\rightarrow
\infty$. (The positive moments of the characteristic polynomials
of random unitary-symmetric matrices were calculated in
\cite{KS1}; for the positive integer moments it was confirmed in
\cite{BH} that, as expected, these results also apply to the large
$N$ limit of matrices in the GOE; see also \cite{MN}.)

Berry \& Keating \cite{BK} (hereinafter referred to as BK) have
recently put forward a general conjecture about the asymptotics of
the negative moments of the characteristic polynomials of random
matrices in the limit as the matrix size tends to infinity when
$\mbox{Im}\mu$ is scaled by the mean eigenvalue density and tends
to zero. This conjecture applies to all negative moments, rather
than just to negative integer moments, and covers all three of the
classical random matrix ensembles (i.e.~the unitary, orthogonal
and symplectic ensembles).  It predicts a highly non-trivial
dependence of the asymptotics on the power to which the polynomial
is raised. This is in contrast to the case when the large-matrix
limit is taken without scaling $\mbox{Im}\mu$ by the mean level
spacing; then the moment asymptotics is much simpler \cite{hko}.

In the case of the Gaussian Unitary Ensemble (GUE) of random
matrices, the conjecture given in BK agrees with the values of the
negative integer moments calculated by Fyodorov in \cite{isint}
and shown to be universal (in the sense that they apply to all
unitary-invariant ensembles of Hermitian matrices) in \cite{sf}.
However, these values also happen to coincide with the
corresponding ones when $\mbox{Im}\mu$ isn't scaled, so this
cannot be said to constitute a test of the non-trivial aspects of
the conjecture.

For the GOE of random matrices the conjecture in BK is that the
ensemble average of $|Z_N(\mu)|^{-k}$ diverges like
$\epsilon^{-\nu(k)}$, as $\epsilon$, $\mbox{Im}\mu$ scaled by the
mean eigenvalue density, tends to zero, with
\begin{equation} \label{bk}
\nu(k)={\rm int}(k)\left( k-\frac{1+{\rm int}(k)}{2}\right).
\end{equation}
It was suggested in BK (page L4) that, in the notation of the
present paper, when $k$ is an integer "it is possible that the
leading-order power-law behaviour (\ref{bk}) is multiplied by a
power of $\log\frac{1}{\epsilon}$".

Our first aim here is to extend the heuristic arguments developed
in BK to recover the logarithmic factor when $k$ takes integral
values; this turns out to be simply $\log\frac{1}{\epsilon}$ for
each $k$. Our second aim is then to prove the resulting expression
by a direct evaluation of the GOE average.  In fact, we are able
to go significantly further in that we calculate the precise
asymptotic form of the moments in the appropriate limit.  The
general expression we obtain (see (\ref{mainres}) and
(\ref{mainint})) takes the form of a multiple integral and is
interesting in its own right, in particular in view of recent
endeavours to understand the analytic structure behind the
so-called replica limit $k\to 0$ \cite{rep1,rep2}.

The heuristic arguments described in BK, which motivate the
conjecture made there, are an application of general techniques
associated with the theory of {\em singularity-dominated strong
fluctuations}.  These techniques have been applied previously to
analyze twinkling starlight \cite{B1}, van Hove-type singularities
\cite{B2}, and the influence of classical periodic orbit
bifurcations on quantum energy level \cite{BKS} and wavefunction
\cite{KP} statistics.  In all of these applications the results
correspond to power-law asymptotics of the moments of fluctuating
quantities as the relevant parameter vanishes, with exponents that
emerge from a competition between different singular
contributions.  It was shown by Hannay \cite{hannay1, hannay2} for
the the moments of the intensity fluctuations beyond a
one-dimensional refracting screen that exactly when one kind of
singularity overtakes another in the competition there is an
additional logarithmic factor. Hannay also obtained the constants
multiplying the various asymptotic contributions in this case.
Importantly, in none of the applications studied previously has it
been possible to prove non-trivial predictions of the theory of
singularity-dominated strong fluctuations by an asymptotic
analysis that could be made rigorous.

In the example we study here, the singularity competition
considered in BK is between clusters of nearly degenerate
eigenvalues.  Clusters involving $p$ eigenvalues give rise to a
contribution to the ensemble average of $|Z_N(\mu)|^{-k}$ that
diverges like $\epsilon^{-\nu_p(k)}$ as $\epsilon \rightarrow 0$.
For a given $k$, the dominating cluster-size is the one for which
the exponent $\nu_p(k)$ is maximal.  It was shown in BK that this
produces the exponent (\ref{bk}).  Here, in Section \ref{sec2}, we
show that for $k$ an integer, when one $p$ takes over from another
as dominant, there is an additional logarithmic factor, as
described above.  In Section \ref{sec3}, we prove this result by
calculating the GOE average explicitly, in the large matrix-size
limit.  This represents the first example where nontrivial
predictions of theory of singularity-dominated strong fluctuations
have been proved.

\section{Cluster contributions}\label{sec2}
We here re-analyze the arguments presented in BK to recover
explicitly the logarithmic factor anticipated there in the case of
negative integer moments of characteristic polynomials of random
matrices in the GOE.

Denoting by $M_p(-k, \epsilon)$ the contribution from clusters of
$p$ eigenvalues (we henceforth refer to this as the $p$-{\em
cluster} contribution) to the GOE average of $|Z_N(\mu)|^{-k}$,
where $\epsilon$ is $\mbox{Im}\mu$ scaled by the mean eigenvalue
density, equations (9) and (10) of BK may be written
\begin{equation}\label{bk9}
M_p(-k,\epsilon)\propto\int_{-X}^{X}dx_1\int_{-X}^{X}dx_2\ldots\int_{-X}^{X}dx_p
\frac{\prod_{m=1}^{p-1}\prod_{n=m+1}^{p}|x_m-x_n|}{\left[(x_1^2+\epsilon^2)
(x_2^2+\epsilon^2)\ldots(x_p^2+\epsilon^2)\right]^{k/2}}.
\end{equation}
To be precise, the limits of integration were given as $-\infty$
and $\infty$ in BK.  This distinction will be important when $k$
is an integer, and not otherwise.  A finite integration range is,
in fact, more appropriate; in the case of the circular ensembles
of random matrix theory because the eigenphases lie in a finite
interval, and in the case of the Gaussian (or similar) ensembles
because the potential effectively limits the range in which the
eigenvalues lie.

Making the change of the variables $x_m=\epsilon u_m$ gives
\begin{equation}\label{bk9a}
M_p(-k,\epsilon)\propto\epsilon^{\frac{p(p+1)}{2}-pk}\int_{-X/\epsilon}^{X/\epsilon}du_1
\int_{-X/\epsilon}^{X/\epsilon}du_2\ldots\int_{-X/\epsilon}^{X/\epsilon}du_p
\frac{\prod_{m=1}^{p-1}\prod_{n=m+1}^{p}|u_m-u_n|}{\left[(u_1^2+1)
(u_2^2+1)\ldots(u_p^2+1)\right]^{k/2}}.
\end{equation}
It was demonstrated in BK that the $p$-cluster contribution
dominates the $k$th moment when $p\le k<p+1$.  It is
straightforward to see that the integral in (\ref{bk9a}) converges
as $\epsilon\rightarrow 0$ in the range $p<k<p+1$.  It is then
asymptotically consistent to replace the limits of integration by
$-\infty$ and $\infty$, and the results of BK hold without change.
When $k$ is an integer, the $p=k$ integral diverges and so must be
treated more carefully.

Let
\begin{equation}\label{bk9b}
I_p(X/\epsilon)=\int_{-X/\epsilon}^{X/\epsilon}du_1
\int_{-X/\epsilon}^{X/\epsilon}du_2\ldots\int_{-X/\epsilon}^{X/\epsilon}du_p
\frac{\prod_{m=1}^{p-1}\prod_{n=m+1}^{p}|u_m-u_n|}{\left[(u_1^2+1)
(u_2^2+1)\ldots(u_p^2+1)\right]^{p/2}}.
\end{equation}
Consider first the case when $p=1$:
\begin{equation}\label{bk9c}
I_1(X/\epsilon)=\int_{-X/\epsilon}^{X/\epsilon}\frac{du_1}{(u_1^2+1)^{1/2}},
\end{equation}
which clearly diverges like $\log\frac{1}{\epsilon}$ as
$\epsilon\rightarrow 0$.

Consider next the case when $p=2$:
\begin{equation}\label{bk9d}
I_2(X/\epsilon)=\int_{-X/\epsilon}^{X/\epsilon}du_1
\int_{-X/\epsilon}^{X/\epsilon}du_2\frac{|u_1-u_2|}{(u_1^2+1)(u_2^2+1)}\propto\int_{-X/\epsilon}^{X/\epsilon}du_1
\int_{-X/\epsilon}^{u_1}du_2\frac{u_1-u_2}{(u_1^2+1)(u_2^2+1)}.
\end{equation}
This can be written as two integrals, one associated with the
first term in the numerator of the integrand and the other
associated with the second term.  It may be seen straightforwardly
that again both integrals diverge like $\log\frac{1}{\epsilon}$ as
$\epsilon\rightarrow 0$.

In the general case
\begin{equation}\label{bk9e}
I_p(X/\epsilon)\propto\int_{-X/\epsilon}^{X/\epsilon}du_1
\int_{-X/\epsilon}^{u_1}du_2\ldots\int_{-X/\epsilon}^{u_{p-1}}du_p
\frac{\prod_{m=1}^{p-1}\prod_{n=m+1}^{p}(u_m-u_n)}{\left[(u_1^2+1)
(u_2^2+1)\ldots(u_p^2+1)\right]^{p/2}}.
\end{equation}
Expanding out the numerator of the integrand, $I_p$ may be
expressed as a sum of integrals, each coming from a term in the
resulting series.  It may be seen immediately that each integral
diverges like $\log\frac{1}{\epsilon}$ as $\epsilon\rightarrow 0$.
Thus when $k$ is an integer the GOE average of $|Z_N(\mu)|^{-k}$
diverges like
\begin{equation}\label{diver}
\epsilon^{-k(k-1)/2}\log\frac{1}{\epsilon}
\end{equation}
as $\epsilon\rightarrow 0$.

\section{GOE negative moments}\label{sec3}

Our purpose now is to prove the result obtained at the end of the
previous section.  We shall do this by making a careful asymptotic
analysis of the exact GOE average defining the moments.

Regularizing the characteristic polynomial $Z_N(\mu)=\det\left(\mu
{\bf 1}_N-\hat{H}\right)$ by taking $\mbox{Im}\mu>0$,
 one may represent negative half-integer powers of the determinant as
a Gaussian integral:
\begin{equation} \label{Gau}
[Z_N(\mu)^{-n/2}] =\frac{1}{(2\pi i)^{nN/2}}\int\prod_{k=1}^n
d{\bf S}_k \exp\left\{\frac{i}{2}\mu\sum_{k=1}^n {\bf S}^{T}_k
{\bf S}_k-\frac{i}{2}\mbox{Tr}\left[\hat{H}\sum_{k=1}^n {\bf S}_k
\otimes{\bf S}^{T}_k\right]\right\},
\end{equation}
where  we have introduced real-valued $N-$dimensional vectors
${\bf S}_k=(s_{k,1},...,s_{k,N})^T$ for $k=1,2,...,n$ so that
$d{\bf S}_k=\prod_{i=1}^N ds_{k,i}$.

Denoting by $\left\langle ...\right\rangle$ the expectation value
with respect to the distribution (\ref{GOE}), our goal is to
calculate the negative integer
 moments
\begin{equation}\label{mom}
{\cal K}_{N,n}^{(1)}(\mu_1)=\left\langle [Z_N(\mu_1)]^{-n/2}
\right\rangle\quad
\end{equation}
as well as the correlation function
\begin{equation}
{\cal K}_{N,n}^{(2)}(\mu_1,\mu_2)=\left\langle
\left[Z_N(\mu_1)Z_N(\mu_2^*)\right]^{-n/2}\right\rangle
\end{equation}
assuming $\mbox{Im}(\mu_1)=\mbox{Im}(\mu_2)>0$. It will be
convenient for us to define $\mu_1=\mu+\frac{\omega}{2}+i\delta$
and $\mu^*_2=\mu-\frac{\omega}{2}-i\delta$, with $\mu,\omega$ and
$\delta$ real and $\delta>0$. Note that when $\omega=0$ the
correlation function reduces to the negative integer moments of
the absolute value of the characteristic polynomial, which are the
main objects of interest here.

We start with (\ref{mom}). Performing the ensemble averaging in
the standard way using the identity
\begin{equation}\label{ident1}
\int d\hat{H}{\cal P}(\hat{H})e^{\pm \frac{i}{2}
\mbox{\small Tr}\left[\hat{H}\hat{A}\right]}=
\exp\left\{-\frac{J^2}{16N}\mbox{Tr}\left[\hat{A}^2+
\hat{A}\hat{A}^T\right]
\right\}
\end{equation}
gives
\begin{equation}\label{avmom1}
{\cal K}_{N,n}^{(1)}(\mu_1)= \frac{1}{(2\pi
i)^{nN/2}}\int\prod_{k=1}^n d{\bf S}_k
\exp\left\{\frac{i}{2}\mu_1\sum_{k=1}^n {\bf S}^{T}_k {\bf
S}_k-\frac{J^2}{8N}\sum_{k,l=1}^n \left({\bf S}^{T}_k{\bf
S}_l\right) \left({\bf S}^{T}_l{\bf S}_k\right)\right\}.
\end{equation}

Introducing an $n\times n$ real symmetric matrix $\hat{Q}$ with
matrix elements $\hat{Q}_{kl}={\bf S}^{T}_k{\bf S}_l$, we note
that the integrand may be conveniently rewritten in the form
\[
\exp\left\{\frac{i}{2}\mu_1 \mbox{Tr} \hat{Q}-\frac{J^2}{8N}
\mbox{Tr}\hat{Q}^2\right\}.
\]
This fact allows us to employ the "integration theorem" proved in
Appendix A of \cite{izhc} and to rewrite the integral in
(\ref{avmom1}) in terms of an integral over the positive definite
matrices $\hat{Q}$:
\begin{equation}\label{k1}
{\cal K}_{N,n}^{(1)}=C_{N,n}^{(1)}\int_{\hat{Q}>0}
 d\hat{Q}e^{-N\left[-i\mu_1\mbox{Tr}{\hat{Q}}+\frac{1}{2}
\mbox{Tr}\hat{Q}^2\right]}\det{\hat{Q}}^{(N-n-1)/2}
\end{equation} provided $N\ge n+1$.
We have also rescaled the integration variable: $\hat{Q}\to
2N\hat{Q}$ so that the overall constant $C^{(1)}_{N,n}$ is given
by
\[
C^{(1)}_{N,n}=\left(-iN\right)^{Nn/2}\pi^{-\frac{n(n-1)}{4}}
\frac{1}{\prod_{j=0}^{n-1}\Gamma\left(\frac{N-j}{2}\right)}
\]
where $\Gamma(z)$ is the Euler gamma-function.

As the last step of the procedure we choose the eigenvalues
$q_1,...,q_n$ and the corresponding eigenvectors of $\hat{Q}$ as
new integration variables. This corresponds to the change of the
volume element
\begin{equation}\label{new1}
d\hat{Q}=\frac{1}{n!}G_n|\Delta\{\hat{q}\}|\prod_{i=1}^n dq_i
d\mu(O_n),
\end{equation}
where $\Delta\{\hat{q}\}= \prod_{i<j}(q_i-q_j)$ is the Vandermonde
determinant and $d\mu(O_n)$ stands for the normalized invariant
measure on the orthogonal group $O(n)$. Here
\begin{equation}\label{new2}
G_n=(\pi)^{\frac{n(n+1)}{4}}
\frac{1}{\prod_{j=1}^{n}\Gamma\left(\frac{j}{2}\right)}
\end{equation}
and the factor $1/n!$ ensures that the integration domain with
respect to all variables $q_k$ can be taken to be $0<q_k<\infty$.

The integrand is obviously $O(n)$ invariant and so we obtain:
\begin{equation}\label{1}
{\cal K}_{N,n}^{(1)}(\mu_1) =\tilde{C}^{(1)}_{N,n}\int_{q_i>0}
\prod_i\left( dq_i e^{iN(\omega/2+i\delta)q_i} q_i^{-(n+1)/2}\right)|\Delta\{\hat{q}\}|
\exp\left\{ -\frac{N}{2}\sum_{i=1}^n A(q_i)\right\},
\end{equation}
where $\tilde{C}^{(1)}_{N,n}=\frac{1}{n!}G_nC^{(1)}_{N,n}$ and
\begin{equation}\label{action}
A(q)=J^2q^2-2i\mu q-\ln{q}.
\end{equation}

We are mainly interested here in the limit of large matrix size,
where one expects the results to show universality. To extract the
leading asymptotics as $N\to \infty$ when $n$ is fixed we employ
the saddle-point method, and consider $N\omega$ as well as $N\delta$
to be of the order unity when $N\to \infty$. The stationary
points of $A(q_i)$ are obviously given by
\begin{equation}\label{s.p.e}
2J^2q_{i}-2i\mu-\frac{1}{q_{i}}=0,
\end{equation}
where $i=1,2,...,n$. Each of these equations has two solutions:
\begin{equation}\label{s.p.s.}
q_{\pm}=\frac{i\mu\pm\sqrt{2J^2-\mu^2}}{2J^2}.
\end{equation}
We would like to choose the spectral parameter $\mu$
to satisfy $|\mu|<J\sqrt{2}$ in accordance with the idea of considering
 the bulk of the spectrum for GOE matrices of large size.
Then only for $q_{+}$ are the real parts positive, and so only in
this case do the corresponding saddle points contribute to the
integral over the positive semiaxis $q>0$. Consequently, among the
$2^{n}$ possible sets of saddle points
$\left(q_{\pm},...,q_{\pm}\right)$ only the choice
\begin{equation}\label{q+}
\hat{q}_+=\mbox{diag}(q_{+},...,q_{+})
\end{equation}
is relevant.

The presence of the Vandermonde determinants makes the integrand
vanish at the saddle-point sets and so care should be taken when
calculating the leading order contribution to the integral. This
turns out to be given by
\begin{equation}\label{sadp}
{\cal K}_{N>>1,n}^{(1)}(\mu_1)=\tilde{C^{(1)}}_{N,n}
(q_{+})^{(N-n-1)/2}e^{-\frac{N}{2}n[J^2q^2_{+}-2i\mu_1q_{+}]}
\int_{-\infty}^{\infty}\prod_{k=1}^n d{\xi_k}
\prod_{k_1<k_2}|\xi_{k_1}-\xi_{k_2}|
e^{-\frac{t}{2}\sum_{k=1}^n\xi^2_k}
\end{equation}
with
\begin{equation}\label{new3}
t=\frac{N(1+2J^2q^2_{+})}{2q^2_{+}}.
\end{equation}
The integral in (\ref{sadp}) is a particular case of the Selberg
integral \cite{Mbook} and can be evaluated explicitly. We do not
give the resulting expression here, because it is not needed for
our purposes.

We note for later purposes that a formula for ${\cal
K}_{N,n}^{(1)}(\mu^*_2)$ can obviously be obtained from the above
expression by taking its complex conjugate and then replacing $\mu^*_1$
with $\mu^*_2$.

We next consider the product of the expression (\ref{Gau}) with
its complex conjugate at a different value of the spectral
parameter and average it over the GOE. From now on we use the
index $\sigma=1,2$ to label the $N$-component vectors ${\bf
S}_{\sigma}$ stemming from the first/second set of integrals. To
write the resulting expression in a compact form it is again
convenient to introduce a $2n\times 2n$ matrix $\hat{Q}$ with the
matrix elements $\hat{Q}^{\sigma_1,\sigma_2}_{kl}= {\bf
S}^{T}_{\sigma_1,k}{\bf S}_{\sigma_2,l}$.  Here $k$ and $l$ take
the values $1,...,n$. In terms of this matrix
\begin{equation}\label{avmom2}
{\cal K}_{N,n}^{(2)}(\mu_1,\mu_2)=\frac{1}{(2\pi)^{Nn}}
\int\prod_{k=1}^n d{\bf S}_{1,k}d {\bf S}_{2,k}
\exp\left\{\frac{i}{2}\mu_1\sum_{k=1}^n {\bf S}^{T}_{1,k} {\bf
S}_{1,k}-\frac{i}{2}\mu_2^*\sum_{k=1}^n {\bf S}^{T}_{2,k} {\bf
S}_{2,k}-\frac{J^2}{8N}\mbox{Tr}\left(\hat{Q}\hat{L}\hat{Q}\hat{L}
\right)\right\},
\end{equation}
where $\hat{L}=\mbox{diag}({\bf 1}_n,-{\bf 1}_n)$. Again employing
the same integration theorem as above and changing $\hat{Q}\to
2N\hat{Q}$ we arrive at
\begin{equation}\label{k2}
{\cal K}_{N,n}^{(2)}(\mu_1,\mu_2)=C_{N,n}^{(2)}\int_{\hat{Q}>0}
 d\hat{Q}e^{-\frac{N}{2}\left[-2i\mbox{Tr}{\hat{M}\hat{Q}}+J^2
\mbox{Tr}\left(\hat{Q}\hat{L}\hat{Q}\hat{L}\right)\right]}
\det{\hat{Q}}^{(N-2n-1)/2},
 \end{equation}
 provided $N\ge 2n+1$, where $\hat{M}=
\mbox{diag}(\mu_1{\bf 1}_n,-\mu_2^* {\bf 1}_n)$ and
\[
C^{(2)}_{N,n}=(N)^{Nn}(\pi)^{-n(2n-1)/2}
\frac{1}{\prod_{j=0}^{2n-1}\Gamma\left(\frac{N-j}{2}\right)}.
\]

This equation differs from its analogue (\ref{k1}) in one
important aspect: it is now of little use to introduce the
eigenvalues/eigenvectors of $\hat{Q}$ as integration variables.
Rather, it is natural to treat $\hat{Q}_L=\hat{Q}\hat{L}$ as a new
matrix to integrate over. Such (non-symmetric!) matrices satisfy
$\hat{Q}_L^{T}=\hat{L}\hat{Q}_L \hat{L}$, have all eigenvalues
real and can be diagonalized by a (pseudo-orthogonal) similarity
transformation $\hat{Q}_L=\hat{T}_0\hat{q}\hat{T}_0^{-1}$, where
$\hat{q}=\mbox{diag}(\hat{q}_1,-\hat{q}_2)$ and the $n\times n$
diagonal matrices $\hat{q}_1,\hat{q}_2$ satisfy
$\hat{q}_1>0\,,\,\hat{q}_2>0$. Pseudo-orthogonal matrices
$\hat{T}_0$ satisfy: $\hat{T}_0^{T}\hat{L}\hat{T}_0=\hat{L}$
and form the group $O(n,n)$ (the corresponding symmetry is
conventionally called a "hyperbolic symmetry" in the random matrix
literature, see \cite{SW}).

It turns out that a  more convenient way to proceed is to
block-diagonalize the matrices $\hat{Q}_L$:
$$
\hat{Q}_L=\hat{T}^{-1}\left(\begin{array}{cc}\hat{P}_1&\\ & -\hat{P}_2
\end{array}\right)\hat{T}\quad,\mbox{where}\quad
\hat{T}\in \frac{O(n,n)}{O(n)\times O(n)}
$$
and $\hat{P}_{1,2}$ are $n\times n$ real symmetric, with positive
eigenvalues $\hat{q}_{1,2}$, respectively. The integration measure
$[d\hat{Q}_L]$ can be derived in terms of the new variables
following the standard steps (see e.g.~\cite{VWZ}) outlined in the
Appendix of the present paper. We arrive at
$[d\hat{Q}]=A\,d\hat{P}_1d\hat{P}_2
\prod_{k_1,k_2}\left(q_{1,k_1}+q_{2,k_2}\right) d\mu(T)$, where
$A=G_n^2/[n!2^{n(n+1)/2}]$ and the last factor is the invariant
measure on the manifold of $T$-matrices. An explicit expression
for it is presented, for reference purposes, in the Appendix.

After all these preparatory steps we arrive at the following
expression:
\begin{eqnarray}\label{k3}
{\cal K}_{N,n}^{(2)}&=&A\,C^{(2)}_{N,n}\int_{\hat{P_1}>0}
\int_{\hat{P_1}>0} d\hat{P}_1d\hat{P}_2\,\, I(\hat{M},\hat{P}_1,\hat{P}_2) \\
\nonumber &\times&
\prod_{k_1,k_2}\left(q_{1,k_1}+q_{2,k_2}\right)
 \det{\left[-\hat{P}_1\hat{P}_2\right]}^{(N-2n-1)/2}
e^{-\frac{NJ^2}{2}\mbox{Tr}\left(\hat{P}_1^2+\hat{P}_2^2\right)},
\end{eqnarray}
where
\begin{eqnarray}\label{coset}
&&I(\hat{M},\hat{P}_1,\hat{P}_2)=\int d\mu(\hat{T})
\exp\left\{iN\mbox{Tr}\left(\begin{array}{cc}
\hat{\mu}_1{\bf 1}_n&\\ & \mu_2^*{\bf 1}_n
\end{array}\right)\hat{T}^{-1}
\left(\begin{array}{cc}\hat{P}_1&\\ & -\hat{P}_2
\end{array}\right)\hat{T}\right\}.
\end{eqnarray}
Employing the explicit parametrization for the matrices $T$ given
in the Appendix we can rewrite the above integral as
\begin{eqnarray}\label{coset1}
&&I(\hat{M},\hat{P}_1,\hat{P}_2)=
e^{iN\frac{\mu_1+\mu_2^*}{2}\sum_k(q_{1k}-q_{2k})}
I_0(\hat{M},\hat{P}_1,\hat{P}_2),\\
&& I_0(\hat{M},\hat{P}_1,\hat{P}_2)=
\int_{-\infty}^{\infty}\prod_{k=1}^n
d\psi_k\prod_{k_1<k_2}|\cosh{\psi_{k_2}}-\cosh{\psi_{k_2}}|\\
\nonumber&&\times \int
[d\mu(O_L)][d\mu(O_R)]\exp\left\{iN\frac{\mu_1-\mu_2^*}{2}\mbox{Tr}
\cosh{\hat{\psi}}\left[\hat{O_L}^T\hat{P_1}\hat{O_L}
+\hat{O_R}^T\hat{P_2}\hat{O_R}\right]\right\},
\end{eqnarray}
where $\hat{O}_{L,R}\in O(n)$, and $\hat{\psi}$ is diagonal.

 In the case of GUE matrices
studied in \cite{isint} a helpful trick under similar conditions
 was to perform the (unitary) group integrals explicitly by employing
the famous Itzykson-Zuber-Harish-Chandra integration formula
\cite{IZHC}. The lack of an analogous formula for the orthogonal
group forces us to take a slightly different route.

It is easy to see that the value of this integral can depend only
on the eigenvalue matrices $\hat{q}_{1}$ and $\hat{q}_{2}$. Let us
therefore introduce the eigenvalues (and corresponding
eigenvectors) of the Hermitian matrices $\hat{P}_{1}>0$ and
$\hat{P}_{2}>0$ as the integration variables. This results in the
following expression:
\begin{eqnarray}\label{2b}
\nonumber {\cal
K}_{N,n}^{(2)}(\mu_1,\mu_2)&=&\tilde{C}^{(2)}_{N,n}
\int_0^{\infty}\prod_i\, dq_{1,i}\,q_{1,i}^{-n-1/2}\,
|\Delta\{\hat{q_1}\}| \int_0^{\infty}\prod_i\, dq_{2,i}\,
q_{2,i}^{-n-1/2}\, |\Delta\{\hat{q_1}\}|
\\
&\times&
\prod_{k_1,k_2}(q_{1,k_1}+q_{2,k_2})\,I(\hat{M},\hat{q}_1,\hat{q}_2)
e^{-\frac{N}{2}\sum_{i=1}^n A(q_{1,i})-\frac{N}{2}\sum_{i=1}^n
A^*(q_{2,i})}
\end{eqnarray}
where
\begin{equation}\label{new4}
\tilde{C}^{(2)}_{N,n}=\frac{1}{2^{n(n+1)/2}n!^3}G_n^4\,C^{(2)}_{N,n},
\end{equation}
\begin{equation}
A(q)=J^{2}q^2-2i\mu q-\ln{q}\quad \mbox{and} \quad
A^*(q)=J^{2}q^2+2i\mu q-\ln{q}.
\end{equation}

Again, we need to perform an asymptotic analysis as $N\to \infty$.
The most interesting regime occurs when one keeps the difference
$\mbox{Re}(\mu_{1}-\mu^*_{2})\equiv \omega$ and the regularization
$\delta$ so small as to ensure $N
\mbox{max}\left(\omega,\delta\right)<\infty$, while $\mu=\mbox{Re}
\frac{(\mu_1+\mu_2)}{2}$ is kept in the range $|\mu|<J\sqrt{2}$.

The stationary points of $A(q)$ and $A^*(q)$
 are now given by
\begin{equation}\label{s.p.e1}
q_{1,i}-i\mu-\frac{1}{q_{1,i}}=0\quad \mbox{and}\quad
q_{2,i}+i\mu-\frac{1}{q_{2,i}}=0,
\end{equation}
where $i=1,2,...,n$. Each of these two equations has two solutions:
\begin{equation}\label{s.p.s.1}
q_{1{\pm}}=\frac{i\mu\pm\sqrt{2J^2-\mu^2}}{2J^2} \quad\mbox{and}\quad
q_{2{\pm}}=\frac{-i\mu\pm\sqrt{2J^2-\mu^2}}{2J^2}\quad,
\end{equation}
but only for $q_{1{+}},q_{2{+}}=q^*_{1{+}}$ are the real parts
positive; that is, only then do the corresponding saddle points
contribute to the integral over the positive semiaxis $q_{1,i}>0$
or $q_{2,i}>0$. Consequently, among the $2^{2n}$ possible sets of
stationary points only the choice
\begin{equation}\label{q+1}
\hat{q_1}=\mbox{diag}(q_{1+},...,q_{1+})\quad,\quad
\hat{q_1}=\mbox{diag}(q^*_{1+},...,q^*_{1+})
\end{equation}
is relevant. This is a major simplification, because for such a
choice the integrand in (\ref{coset1}) turns out to be independent
of the matrices $\hat{O}_1,\hat{O}_2$.

Taking care of the Vandermonde determinants when calculating the
fluctuations around the chosen saddle points and remembering that
\begin{equation}
q_1+q_1^*=\pi\rho(\mu)\quad,\quad q_1q_1^*=1/2J^2
\end{equation}
where $\rho(\mu)=\frac{1}{\pi J^2}\sqrt{2J^2-\mu^2}$ is the mean
density of eigenvalues for GOE matrices, we observe that when the
asymptotic expression for the correlation function under
consideration is divided by the product of the negative moments
(\ref{sadp}) the Selberg integrals cancel out, as well as all of
the exponential factors too. The resulting expression amounts to
\begin{equation}\label{mainres}
{\cal K}_n(\mu_1,\mu_2)=\lim_{N\to \infty} \frac{\left\langle
\left[\det{(\mu_1 {\bf 1}_N-\hat{H})} \det{(\mu^*_2 {\bf
1}_N-\hat{H})}\right]^{-n/2}\right\rangle} {\left\langle
\det{(\mu_1 {\bf 1}_N-\hat{H})}^{-n/2}\right\rangle
\left\langle\det{(\mu^*_2 {\bf 1}_N-\hat{H})}^{-n/2}\right\rangle}
={\cal C}\times F^{GOE}_n\left(\epsilon\right),
\end{equation}
where
\begin{equation}
 F^{GOE}_n(\epsilon)=e^{n\epsilon}\int_1^{\infty}
\prod_{k=1}^n
\frac{d\lambda_k}{\sqrt{\lambda_k^2-1}}\,\prod_{k_1<k_2}
|\lambda_{k_1}-\lambda_{k_2}|e^{-\epsilon\sum_{k=1}^n\lambda_k},
\label{mainint}
\end{equation}
in which we have introduced variables
$\lambda_k=\cosh{\psi_k}\in[1,\infty)$,
\begin{equation} \label{new15}
\epsilon=-iN\pi\rho(\mu)(\mu_1-\mu_2^*)/2,
\end{equation}
and
\begin{equation}\label{new16}
 {\cal C}=(\pi\rho J)^{n^2}\left(\frac{N}{2}\right)^{n^2/2}
\frac{(2\pi)^{n/2}}{n!}
\frac{1}{\left[\prod_{j=1}^n\Gamma\left(\frac{j}{2}\right)\right]^2}.
\end{equation}
This expression is valid for all $|\epsilon|<\infty$,
i.e. as far as $(\mu_1-\mu_2^*)=O(1/N)$ and constitutes
 one of the main results of the present paper. In the subsequent
analysis we concentrate on the moments of characteristic polynomials
and thus treat $\epsilon$ as a real parameter.

It is instructive to compare (\ref{mainint}) with its counterpart
for the Gaussian Unitary Ensemble (see \cite{VZ}; in \cite{isint}
the corresponding expression is implicit):
\begin{eqnarray}\label{GUE}
F^{GUE}_n(\epsilon)=e^{n\epsilon}\int_1^{\infty} \prod_{k=1}^n
d\lambda_k\,\prod_{k_1<k_2} (\lambda_{k_1}-\lambda_{k_2})^2
e^{-\epsilon\sum_{k=1}^n\lambda_k}.
\end{eqnarray}
The latter integral is a specific case of the Selberg integral
\cite{Mbook} and can be immediately evaluated, yielding
\begin{eqnarray}\label{GUE1}
F^{GUE}_n(\epsilon)= \frac{1}{\epsilon^{n^2}}
\prod_{j=0}^{n-1}j!(j+1)!.
\end{eqnarray}
Such a formula exemplifies a `normal' dependence of the negative
moments on $\epsilon$: namely, one can extract the rate of
divergence as $\epsilon\to 0$ by analysing the perturbative
expansion of the integral as $\epsilon\to\infty$.
Performing the latter limit is effectively the same as considering
the case when $\delta=\mbox{Im}\mu_{1,2}$ is left unscaled by the mean
eigenvalue density (see the Introduction). In other words, it is
equivalent to considering the limit $\delta\to 0$ {\it after}
taking $N\to \infty$. Thus for the GUE the asymptotics of the
negative moments is the same irrespective of the order in which
limits are taken, and so is relatively uninteresting.

The integral (\ref{mainint}) behaves in this sense `anomalously'.
It does not belong to the class of Selberg integrals and apart
from when $n=1$ (in which case it just yields the Macdonald
function $K_0(\epsilon)$) we have failed to evaluate it explicitly
in a simple closed form. We therefore proceed to analyze the
limits $\epsilon\to \infty$ and $\epsilon\to 0$ separately.

In the perturbative region $\epsilon>>1$ the integral is obviously
dominated by a small vicinity of the lower limit:
$\lambda_k-1<<1$. Introducing variables $x_k\in[0,\infty)$ such
that $\lambda_k=1+x_k/\epsilon$, we immediately see that
asymptotically the integral is again of Selberg type:
\begin{eqnarray}\label{GOE1}
&& F^{GOE}_n(\epsilon>>1)=\frac{1}{2^{n/2}\epsilon^{n^2/2}}
\int_0^{\infty} \prod_{k=1}^n
\frac{dx_k}{\sqrt{x_k}}\,\prod_{k_1<k_2}
|x_{k_1}-x_{k_2}|e^{-\sum_{k=1}^n x_k}
\\
&&=\frac{1}{2^{n/2}\epsilon^{n^2/2}}
\prod_{j=0}^{n-1}\frac{\Gamma\left(\frac{3+j}{2}\right)
\Gamma\left(\frac{1+j}{2}\right)}{\Gamma\left(\frac{3}{2}\right)}
=\frac{1}{\epsilon^{n^2/2}}\frac{n!}{(2\pi)^{n/2}}
\left[\prod_{j=1}^{n}\Gamma\left(\frac{j}{2}\right)\right]^2.
\end{eqnarray}
We then see that the perturbative behaviour for GOE moments is essentially
of the same type as that for GUE moments:
\begin{equation}\label{pert}
{\cal K}_n(\mu_1,\mu_2)=
\left(\pi\rho(\mu)J\right)^{n^2}\left(\frac{N}{2\epsilon}\right)^{n^2/2}=
\left(\frac{\pi\rho(\mu)J}{-i[\mu_1-\mu_2^*]/J}\right)^{n^2/2}
\end{equation}

In contrast to this, in the non-perturbative region $\epsilon\to
0$ the behaviour of the GUE and GOE moments is very different. In
this limit the integral is dominated by
$\lambda_k\sim\epsilon^{-1}>>1$ and it is natural to introduce
rescaled variables $y_k=\epsilon \lambda_k$, leading to
\begin{eqnarray}\label{GOE2}
 F^{GOE}_n(\epsilon<<1)=\frac{1}{\epsilon^{n(n-1)/2}}
 \int_{\epsilon}^{\infty}\ldots\int_{\epsilon}^{\infty}
 \prod_{k=1}^n
\frac{dy_k}{y_k}\,\prod_{k_1<k_2}
|y_{k_1}-y_{k_2}|e^{-\sum_{k=1}^n y_k}.
\end{eqnarray}
Note that one cannot set the lower limit of integration with
respect to the variables $y_k$ to be zero, because the
corresponding integrals diverge logarithmically there. To extract
the leading order behaviour as $\epsilon\to 0$ we differentiate
the function $\tilde{F_n}(\epsilon)=\epsilon^{n(n-1)/2}
\,F^{GOE}_n(\epsilon<<1)$ with respect to its argument, reducing
it asymptotically to a Selberg-type integral
\begin{eqnarray}
&&\frac{d}{d\epsilon}\tilde{F_n}(\epsilon)=
-\frac{n}{\epsilon}e^{-\epsilon}
\int_{\epsilon}^{\infty}\frac{dy_2}{y_2}e^{-y_2}
\ldots \int_{\epsilon}^{\infty}\frac{dy_n}{y_n}e^{-y_n}
(y_2-\epsilon)...(y_n-\epsilon)\prod_{2\le k<l\le n}|y_k-y_l|
\\ && {\longrightarrow}
-\frac{n}{\epsilon}\int_{0}^{\infty}\prod_{k=1}^{n-1} dy_k
e^{-y_k}\times \prod_{1\le k<l\le
(n-1)}|y_k-y_l|=-\frac{n}{\epsilon}
\prod_{j=0}^{n-2}\frac{\Gamma\left(\frac{3+j}{2}\right)
\Gamma\left(\frac{2+j}{2}\right)}{\Gamma\left(\frac{3}{2}\right)}.
\end{eqnarray}
Thus, we conclude that for all integer $n\ge 1$
\begin{equation}\label{final}
F^{GOE}_n(\epsilon\to 0)=\frac{2^{n-1}}{\pi^{n/2}}
\,n\prod_{j=0}^{n-1}
\left[\Gamma\left(1+\frac{j}{2}\right)\right]\left[\Gamma
\left(\frac{j+1}{2}\right)\right]\,\frac{\ln{1/\epsilon}}
{\epsilon^{n(n-1)/2}}.
\end{equation}
This anomalous behaviour parametrically agrees with that predicted
by the heuristic theory of dominating singularities outlined in
Section \ref{sec2}; see in particular (\ref{diver}).

\section*{Acknowledgements}
This research was supported by a Brunel University Vice Chancellor
grant as well as by EPSRC grant Gr/13838/01 "Random Matrices close
to unitary or Hermitian".

\vskip 1cm

\section*{Appendix - Calculation of the Jacobian}

To evaluate the integral in (\ref{k2}) one needs to calculate the
Jacobian generated by the variable transformation
$\hat{Q}_L=\hat{T}^{-1}\hat{P}\hat{T}$, with $\hat{P}=
\mbox{diag}(\hat{P}_1,-\hat{P_2})=\hat{P}^T$ and
$\hat{P}_{1}>0,\hat{P}_{2}>0$ being real symmetric $n\times n$
matrices. For the matrices $\hat{T}\in \frac{O(n,n)}{O(n)\times
O(n)}$ we employ the following explicit parametrisation in terms
of a real $n\times n$ matrix $\hat{t}$:
\[
\hat{T}=\left(\begin{array}{cc} \sqrt{1+\hat{t}\hat{t}^T}&
\hat{t}\\ \hat{t}^T& \sqrt{1+\hat{t}^T\hat{t}}\end{array}\right)
\quad \mbox{hence} \quad \hat{T}^{-1}=\left(\begin{array}{cc}
\sqrt{1+\hat{t}\hat{t}^T}& -\hat{t}\\ -\hat{t}^T&
\sqrt{1+\hat{t}^T\hat{t}}\end{array}\right).
\]

It is convenient to follow the scheme suggested in \cite{VWZ}. One
starts by considering the relation between the matrix
differentials:
\begin{equation}
d\hat{Q}=\hat{T}^{-1}d\hat{\tilde{Q}}\hat{T}\quad,\quad
d\hat{\tilde{Q}}=d\hat{P}+\left(\hat{P}d\hat{\tau}-
d\hat{\tau}\hat{P}\right),
\end{equation}
where we have introduced the notation
$d\hat{\tau}=d\hat{T}\hat{T}^{-1}$. To calculate the Jacobian the
difference between $d\hat{Q}$ and $d\hat{\tilde{Q}}$ is immaterial
and we omit the tilde henceforth. Partitioning the matrix
$d\hat{Q}$ into four $n\times n$ sub-blocks $d\hat{q}_{pq}$,
$p,q=1,2$, one then rewrites the above relation blockwise:
\[
d\hat{q}_{11}=d\hat{P}_1+\left(\hat{P}_1d\hat{\tau}_{11}-
d\hat{\tau}_{11}\hat{P}_1\right)\quad,\quad
d\hat{q}_{22}=-d\hat{P}_2-\left(\hat{P}_2d\hat{\tau}_{22}-
d\hat{\tau}_{22}\hat{P}_2\right)
\]

\[
d\hat{q}_{12}=\hat{P}_1\,d\hat{\tau}_{12}+
d\hat{\tau}_{12}\hat{P}_2 \quad,\quad
d\hat{q}_{21}=d\hat{q}^T_{12}.
\]

Inspecting the block structure of the corresponding Jacobian,
symbolically written as
$J=\mbox{det}\left(d\left[\hat{q}_{11},\hat{q}_{22},
\hat{q}_{12}\right]/d\left[\hat{P}_1,\hat{P}_2,\hat{\tau}
\right]\right)$, one may easily verify that
\[
J=\mbox{det}\left(d\left[
\hat{q}_{12}\right]/d\left[\hat{\tau}_{12}\right]\right)=
\mbox{det}\left(\hat{P}_1\otimes {\bf 1}_n+{\bf 1}_n\otimes
\hat{P}_2 \right)=\prod^n_{i,j}\left(q_{1,i}+q_{2,j}\right),
\]
where $q_{1,i}$ and $q_{2,i}$ are (positive) eigenvalues of
the matrices $\hat{P}_{1},\hat{P}_{2}$.
Then an intermediate result for the measure can be
schematically written as
\[
d\hat{Q}=\prod^n_{i,j}\left(q_{1,i}+q_{2,j}\right)
\,d\hat{P}_1d\hat{P}_2\mbox{det}\left(d\left[
\hat{\tau}_{12}\right]/d\left[\hat{t}\right]\right)\,d\hat{t},
\]
where, explicitly,
\[
-d\left[\hat{\tau}_{12}\right]=\left[dT_{11}\right]\left[T^{-1}\right]_{12}+
[dT]_{12}\left[T^{-1}\right]_{22}=
d\left[\sqrt{1+\hat{t}\hat{t}^T}\right]\, \hat{t} +
d\hat{t}\,\sqrt{1+\hat{t}^T\hat{t}}.
\]

To calculate the remaining determinant we employ the singular
value decomposition
$\hat{t}=\hat{O}_L^{-1}\sinh{\hat{\theta}}\hat{O}_R$ expressing
$\hat{t}$ in terms of the two real orthogonal $n\times n$ matrices
$\hat{O}_{L,R}\in O(n)$ and a real diagonal matrix
$\hat{\theta}=\mbox{diag} \left(\theta_1,...,\theta_n\right)$,
assuming, for uniqueness, $\theta_1>...>\theta_n$. Then
$\sqrt{1+\hat{t}\hat{t}^T}=
\hat{O}_L^{-1}\cosh{\hat{\theta}}\hat{O}_L$ and
$\sqrt{1+\hat{t}^T\hat{t}}=
\hat{O}_R^{-1}\cosh{\hat{\theta}}\hat{O}_R$. Further introducing
$d\hat{v}_{L,R}=d\hat{O}_{L,R}\,\left[\hat{O}_{L,R}\right]^{-1}$
and
$d\hat{\tilde{\tau}}=\hat{O}_Ld\left[\hat{\tau}_{12}\right]\hat{O}^{-1}_R$
we find, after straightforward manipulations,
\[
d\hat{\tilde{\tau}}=d\hat{\theta}+\sinh{\hat{\theta}}d\hat{v}_R
\cosh{\hat{\theta}}-\cosh{\hat{\theta}}d\hat{v}_L\sinh{\hat{\theta}}.
\]
Next, differentiating $\hat{O}_{L,R}\hat{O}_{L,R}^T=1$, we observe
that $d\hat{v}_{L,R}$ must be antisymmetric, hence $d\hat{v}_{ii}
=0$ and $d\hat{v}_{j\ne i}=-d\hat{v}_{ij}$, from which it is clear
that $ [d\hat{\tilde{\tau}}]_{ii}=\theta_i$ for all $i=1,...,n$.
At the same time, for any of the $n(n-1)/2$ pairs $1\le i<j\le n$
we have, in vector notation, the relation between the
differentials
\begin{equation}
\left(\begin{array}{c} d\tilde{\tau}_{ij} \\ d\tilde{\tau}_{ji}\end{array} \right)
=\left(\begin{array}{cc}\sinh{\theta_i}\cosh{\theta_j}&
-\cosh{\theta_i}\sinh{\theta_j}\\-\cosh{\theta_i}\sinh{\theta_j}&
\sinh{\theta_i}\cosh{\theta_j}\end{array}\right)\,\,
\left(\begin{array}{c} (d\hat{v}_R)_{ij}\\
 (d\hat{v}_L)_{ij}\end{array}\right).
\end{equation}
The Jacobian in question then reduces to a product of the
determinants of the matrices entering in the above equation, which
are simply $|\sinh^2{\theta_i}-\sinh^2{\theta_j}|=
|\cosh{2\theta_i}-\cosh{2\theta_j}|/2$. Finally, we introduce
$\psi_i=2\theta_i$ , remove the relative ordering of $\psi_i$ in
favour of the factor $1/n!$ in the measure, and remember that
$[d\hat{v}_{L,R}]$ gives rise to the product of invariant measures
$[d\mu(O_{L,R})]$ on the orthogonal group $O(n)$ (which we assumed
to be normalized to unity). The measure $[d\hat{Q}_L]$ in the
coordinates $\hat{P}_{1,2}, \hat{\psi},\hat{O}_{L,R}$ then assumes
the following form:
\begin{equation}\label{measure}
[d\hat{Q}]=\frac{G^2_n}{2^{n(n+1)/2}\,n!}\prod^n_{i,j}
\left(q_{1,i}+q_{2,j}\right)
\prod_{1\le i<j\le n}|\cosh{\psi_i}-\cosh{\psi_j}|
\,d\hat{P}_1d\hat{P}_2d\hat{\psi}\,[d\mu(O_{L})]\,[d\mu(O_R)]
\end{equation}
where $-\infty\le \psi_i<\infty$ for $i=1,...n$.

To conclude, we give the explicit expression for the following
combination used in the main text:
\begin{eqnarray}
\nonumber && \mbox{Tr}\left[\left(\begin{array}{cc}
\hat{\mu}_1{\bf 1}_n&\\ & \mu_2^*{\bf 1}_n
\end{array}\right)\hat{T}^{-1}
\left(\begin{array}{cc}\hat{P}_1&\\ & -\hat{P}_2
\end{array}\right)\hat{T}\right]\\ \nonumber
&=&\mbox{Tr}\left(\begin{array}{cc}
\hat{\mu}_1{\bf 1}_n&\\ & \mu_2^*{\bf 1}_n
\end{array}\right)
\left(\begin{array}{cc}\cosh{\hat{\psi}/2} & -\sinh{\hat{\psi}/2}
\\ -\sinh{\hat{\psi}/2} & \cosh{\hat{\psi}/2}
\end{array}\right)
\left(\begin{array}{cc}\hat{P}_L&\\ & -\hat{P}_R
\end{array}\right)
\left(\begin{array}{cc}\cosh{\hat{\psi}/2}& \sinh{\hat{\psi}/2}
\\ \sinh{\hat{\psi}/2} & \cosh{\hat{\psi}/2}
\end{array}\right)\\
\nonumber
&=&\mbox{Tr}\left[\hat{P}_L
\left(\mu_1\cosh^2{\hat{\psi}/2}
-\mu_2^*\sinh^2{\hat{\psi}/2}\right)\right]
-\mbox{Tr}\left[\hat{P}_R\left(\mu^*\cosh^2{\hat{\psi}/2}-
\mu\sinh^2{\hat{\psi}/2}\right)\right]
\\
&=&\frac{1}{2}(\mu_1+\mu_2^*)\mbox{Tr}\left(\hat{P}_L-\hat{P}_R\right)
+\frac{1}{2}(\mu_1-\mu_2^*)\left(\mbox{Tr}\hat{P}_L\cosh{\hat{\psi}}
+\mbox{Tr}\hat{P}_R\cosh{\hat{\psi}}\right)
\end{eqnarray}
where we have introduced matrices
$\hat{P}_L=\hat{O}_L\hat{P}_1\hat{O}^{-1}_L$ and
$\hat{P}_R=\hat{O}_R\hat{P}_2\hat{O}^{-1}_R$ having the same
eigenvalues $\hat{q}_1=\mbox{diag}(q_{1,1},...,q_{1,n})$ and
$\hat{q}_2=\mbox{diag}(q_{2,1},...,q_{2,n})$ as the matrices
$\hat{P}_{1,2}$.

\end{document}